\documentclass[prb,twocolumn,superscriptaddress]{revtex4}
\usepackage{amsfonts}
\usepackage{amssymb}
\usepackage{amsmath}
\usepackage{bm}
\usepackage{color}
\usepackage{graphicx}
\usepackage{time}
\usepackage[colorinlistoftodos]{todonotes}
\usepackage{sidecap}
\newcommand{\bfi}[1]{\mbox{\boldmath $#1$}}

\begin{document}
%
%
\title{Deformed Explicitly Correlated Gaussians}

\author{Matthew Beutel}
\affiliation{Department of Physics and Astronomy, Vanderbilt University, Nashville, Tennessee, 37235, USA}
\author{Alexander Ahrens}
\affiliation{Department of Physics and Astronomy, Vanderbilt University, Nashville, Tennessee, 37235, USA}
\author{Chenhang Huang}
\affiliation{Department of Physics and Astronomy, Vanderbilt University, Nashville, Tennessee, 37235, USA}

\author{Yasuyuki Suzuki}
\affiliation{Department of Physics, Niigata University, Niigata, Japan}
\author{K\'alm\'an Varga}
\email{kalman.varga@vanderbilt.edu}
\affiliation{Department of Physics and Astronomy, Vanderbilt University, Nashville, Tennessee, 37235, USA}
\begin{abstract}
Deformed correlated Gaussian basis functions are introduced and their matrix elements are calculated.
These basis functions can be used to solve problems with nonspherical
potentials. One example of such potential is the dipole self-interaction term in the Pauli-Fierz Hamiltonian. Examples are
presented showing the accuracy and necessity of deformed Gaussian basis
functions to accurately solve light-matter coupled systems in cavity QED.
\end{abstract}
\maketitle
\section{Introduction}
Few-body approaches have reached very high accuracy in atomic and
molecular systems \cite{boys60,singer60,kolos63a,drake70a,drake91a,yan97a,
korobov00a,nakatsuji07a,ryzhikh97b,PhysRevA.84.012509,bubin04a,stanke06a,
cencek95b,1.4794192,1.4826450,1.4834596,cr200419d,1.4897634,1.4873916} 
and these calculations proved to be indispensible explaining  properties such as the 
electron correlations\cite{lin83a}, relativistic effects\cite{PhysRevA.84.012509,cencek96a}, 
molecular bonds\cite{richard94a,strasburger99a,cencek00a}, 
and quantum dynamics of nuclei\cite{stanke07d,bubin07b,pachucki09a,holka11a}.
As an example, the accuracy of theoretical prediction 
\cite{PhysRevLett.122.103003} and experimental measurement
\cite{PhysRevLett.122.103002} has reached the level of 1 MHz for 
the dissociation energy of the H$_2$ molecule. The accuracy allows
benchmarking the theory against measurement to answer
fundamental questions (e.g. about the nature of the physical
constants) and the development of  accurate approximations 
for efficient computational approaches.

\indent Not all few-body approaches are created equal. There are approaches
with distinct advantages for certain systems and well-known limitations
for others. Hylleraas-type wave functions work extremely well in  two- and three-electron 
atomic systems,\cite{hylleraas28a,hylleraas29a,PhysRevA.87.042518,bhatia67a,yan97a,ho99a,
korobov96a,korobov00a,korobov02a,korobov06b,yan08a,wang11a} but
the extension of the Hylleraas basis approach beyond three
electrons is very tedious because the analytical calculation of the
matrix elements \cite{king11a} is difficult. Hyperspherical approaches \cite{PhysRevA.86.062513,PhysRevA.10.1986,PhysRevA.82.022706,			
NIELSEN2001373,PhysRevA.82.022706,LIN19951,PhysRevA.52.3362,PhysRevA.63.062705}
have also been succesfully used but have similar limitations although
there are new approaches to circumvent the size restrictions \cite{PhysRevA.80.022504,10.1093/ptep/ptv052,
Suzuki2018,PhysRevA.89.012503,PhysRevC.103.024004,PhysRevC.101.014002}.

The matrix elements of Explicitly Correlated Gaussian (ECG) can be
calculated for any number of particles. Because of this,  ECGs became very popular tools in high accuracy
calculations \cite{PhysRevA.102.062825,PhysRevLett.118.043001,
PhysRevA.100.032504,1.4826450,
B211193D,4890373,varga_1998,3491029,1.4834596,cr200419d,PhysRevLett.113.213201,
1.4897634,RevModPhys.85.693,PhysRevA.92.013608,PhysRevA.89.052501,PhysRevA.84.012509,
PhysRevLett.113.073004,PhysRevA.89.032510,PhysRevA.92.012513,1.4873916,4731696,
PhysRevA.80.062510,ZaklamaTimothy2020MEoO,ctx19577663200003276,KEDZIORSKI2020137476,
ctx19578016840003276,MATYUSEdit2013OtCo,doi:10.1063/5.0051237,PhysRevA.102.052806,cr200419d,PhysRevA.102.022803,PhysRevE.101.023313,NASIRI2020143,PhysRevA.100.042503,stanke_2019,M_ller_2019,
PhysRevA.99.012504,doi:10.1063/1.5050462,doi:10.1063/1.5009465,ADAMOWICZ201787,PhysRevA.95.062509,
Fedorov2016,doi:10.1063/1.4948708,BUBIN2016122,PhysRevA.92.062501,PhysRevA.89.012506,PhysRevLett.111.193401,
PhysRevA.87.054501,PhysRevD.103.074503}. The practice of using ECGs as basis functions has been around since 1960 
\cite{boys60,singer60}. The quadratic form involving inter-particle distances in ECGs 
permits the reduction of the Hamiltonian matrix elements to very simple analytic 
expressions and the algebraic complexity of the matrix elements does not 
change with the number of particles.  Further, the matrix elements can be generalized 
for an arbitrary angular momentum
\cite{doi:10.1063/1.4948708,stanke_2019,4731696,doi:10.1063/1.4948708,sharkey_pra_80_062510_2009,Varga1998,4890373}.
These matrix elements depend on the Gaussian parameters of the ECGs 
which should be carefully
optimized\cite{kozlowski92b,suzuki98a,komasa95,PhysRevA.99.012504,bubin10a,tung11a,sharkey11c,bubin2008}
to get highly accurate variational upper bounds. The Gaussian parameters are 
most often chosen to be real, but the extension to complex parameters has
also been tested \cite{bubin06b,PhysRevA.95.062509,BUBIN2016122}.
Systems with periodic boundary conditions have also been investigated
\cite{PhysRevA.87.063609}.

The wide range of applications of ECGs has been demonstrated in 
several recent reviews\cite{cr200419d,B211193D,RevModPhys.85.693}. Currently, 
high accuracy ECG calculations are actively pursued for relatively
large systems, e.g. five-body calculations of  the energy of the H$_3^+$ \cite{ctx19577663200003276}, 
or the Beryllium atom with finite nuclei mass
\cite{PhysRevA.100.032504}, or a six-particle calculation of the 
Boron atom \cite{PhysRevLett.118.043001} and the  singly charged Carbon ion
\cite{PhysRevA.102.062825}. These calculations reached high accuracy,
and using relativistic corrections are comparable to the experimental
data. To reach this accuracy for large systems one needs a large basis
dimension. For example, in Ref. \cite{PhysRevA.100.032504}, 16000 basis
functions were used. Additionally, ECGs have been applied to the nuclei 
in multi-cluster approximations 
\cite{suzuki2008,PhysRevC.89.064303,PhysRevC.100.024334,
Suzuki2021,10.1093/ptep/pts015,AoyamaS2012FSwa,suzuki2008}. While these cases do not reach the same level of accuracy as the atomic and molecular cases, 
ECGs offer the unique advantage of treating the 
nuclear dynamics in an
efficient way. 

The  ECGs are not restricted to bound state problems. More
recently scattering of composite particles have also been studied
using ECG's combined with the confined variational method 
\cite{PhysRevA.103.052803,PhysRevLett.101.123201,PhysRevA.103.022817,PhysRevA.103.042814,PhysRevA.101.042705,
PhysRevA.100.032701,PhysRevLett.103.223202,PhysRevA.78.042705}.

There are many works that have evaluated the matrix elements of ECG’s for spherical 
(L = 0) cases \cite{suzuki98a,kozlowski91a,kozlowski92a,cencek93a,cencek95a,varga95,bubin2008,
PhysRevD.103.074503,PhysRevA.87.063609}. Spherical ECGs have been used in a variety of applications, 
such as the study of Efimov physics\cite{PhysRevLett.113.213201},
hyperfine splitting \cite{PhysRevLett.111.243001,PhysRevA.89.032510},
quantum electrodynamic corrections\cite{PhysRevA.92.012513},
Fermi gases of cold atoms\cite{PhysRevA.92.013608}, and 
potential energy curves \cite{3491029}.

There are many systems where nonspherical (L $>$ 0)
ECGs are necessary (e.g. polyatomic molecules or excited states of
atoms), but the calculation of the matrix elements
of these functions is more complicated.  There are two different ways
that have been proposed. In the first one the Gaussian centers are shifted, which introduces nonspherical components into the basis functions
\cite{PhysRevA.102.022803,doi:10.1063/1.5050462,suzuki98a,4890373,doi:10.1063/1.5009465,
doi:10.1080/00268976.2013.783938}.
The advantage of this approach is that the calculation of the matrix
elements remain simple, and the disadvantage is that the desired angular
momentum has to be built-in \cite{4890373,PhysRevA.99.052512,PhysRevA.102.052806}
or has to be projected out \cite{PhysRevA.102.022803,doi:10.1063/1.5050462}. 

The second possibility is to multiply the ECGs with polynomials of the interparticle coordinates.
Different approaches have been developed to calculate the
matrix elements in this case. For example, one can restrict the calculation
for a special $L$ value, and explicitly work out the formalism for
that case. For example, Refs. \cite{PhysRevA.80.062510,sharkey11b,sharkey11c,sharkey11d,sharkey_jcp_132_184106_2010} calculated the energy and
energy gradient matrix elements for $L = 1$, while Ref. \cite{sharkey11b} tackled
D states. Alternatively, representations using 
``global vectors'' have been put forward\cite{suzuki98a,varga95,varga_1998,SuzukiY1998Ndoo} and this approach has been
developed further \cite{suzuki2008,4731696,MATYUSEdit2013OtCo}. In the global vector
representation, a vector $\mathbf{v}$, formed as a linear combination of all particle coordinates, 
is used as an argument of spherical harmonics to define the orbital momentum. The
coefficients in the linear combination are treated as real-valued
variational parameters. The advantage of this approach is that
the calculation of the matrix elements remains relatively
simple. The disadvantage is that the optimization of the
variational parameters is difficult, and not all possible partial
wave expansion components can be readily represented. 

Another approach to calculating matrix elements of non-spherical ECGs
is to calculate  the matrix elements of 
the 1D case analytically\cite{ZaklamaTimothy2020MEoO}, and then 
generalize to 2D or 3D using tensor products  if needed. The
advantage of this approach is that the ECGs parameters can be different
in different directions and problems with nonspherical potentials can
be solved. 

Finally, direct calculation of nonspherical ECG matrix elements for a
general case has also been worked out \cite{doi:10.1063/1.4948708}. In
this case, the matrix elements are calculated for any desired product of
single-particle coordinates.

The goal of this paper is to introduce Deformed Explicitly Correlated Gaussians (DECGs).
In the DECGs, the Gaussian parameters are different in the $x,y$, and
$z$ directions. In addition, we will also allow the Gaussian centers
to be placed in an arbitrary position and use the position as a
variational parameter. The DECGs can be used to solve problems where the
potential is non-spherically symmetric.

The introcution of DECGs is motivated by the recent interest in
light-matter coupled systems, particularly atoms and molecules in
cavity  QED \cite{FriskKockum2019,doi:10.1021/acsphotonics.7b01279,PhysRevA.98.043801,
QEDHAM,rokaj2021free,doi:10.1021/acs.jpclett.0c01556,10.21468/SciPostPhys.9.5.066,
PhysRevLett.121.113002,PhysRevLett.122.193603,doi:10.1021/acs.jpclett.8b02609,
PhysRevLett.126.153603,Schafer4883,Ruggenthaler2018,Flick15285,Flick3026,https://doi.org/10.1002/qute.201900140,
doi:10.1063/5.0012723,PhysRevB.98.235123}. The light-matter coupled systems are usually
described on the level of the Pauli-Fierz (PF) nonrelativistic QED
Hamiltonian \cite{QEDHAM,Ruggenthaler2018,acs.jpcb.0c03227}. In this
Hamiltonian, there is a dipole self-interaction term,
$(\vec{\lambda}\cdot\vec{D})^2$, where $\vec{\lambda}$ is the coupling
vector of the photons and $\vec{D}$ is the dipole moment of the system. 
This term introduces a nonspherical potential into the Hamiltonian which
makes the calculation difficult. The situation is somewhat
reminiscent of the magnetic Hamiltonian where the potential is
cylindrically symmetric and Gaussians have to be tailored to this
symmetry \cite{PhysRevA.89.052501,PhysRevA.92.033401}. 

The outline of this paper is as follows. Succeeding the introduction we will introduce our notation and formalism, while Secs. \ref{A}-\ref{F} 
will provide the calculations for the overlap matrix, dipole self-interaction removal, electron-photon coupling in addition to the kinetic and potential energy operators. Numerical examples are given in Sect. \ref{III}. To make the paper 
more easily readable, useful but not essential equations are collected in the Appendices. Atomic units are used in the paper. 
\vskip 0.5cm

\section{Formalism}
We consider a system of $N$ particles with positions 
$\vec{r}_1,...\vec{r}_N$, where $\vec{r}_i=(x_i,y_i,z_i)$, and charges $q_1,...,q_N$. 
We define
\begin{equation}
    {\bfi x}=\left(
\begin{array}{c}
x_1\\
x_2\\
\vdots \\
x_N
\end{array}
\right)
\label{vec}
\end{equation}
and $\bfi y$ and $\bfi z$ similarly. We also define 
\begin{widetext}
\begin{equation}
    \vec{\bfi r}=\left(
\begin{array}{c}
{\bfi x}\\
{\bfi y}\\
{\bfi z}
\end{array}
\right)=
\left(
\begin{array}{c}
x_1\\
x_2\\
\vdots \\
x_N\\
y_{1}\\
\vdots \\
y_{N}\\
z_{1}\\
\vdots\\
z_{N}
\end{array}
\right)=
\left(
\begin{array}{c}
r_1\\
r_2\\
\vdots \\
r_N\\
r_{N+1}\\
\vdots \\
r_{2N}\\
r_{2N+1}\\
\vdots\\
r_{3N}
\end{array}
\right)
.
\end{equation}
\end{widetext}
So in the following $\vec{a}$ will be used for 3-dimensional vectors, and ${\bfi a}$ will be used for a set of single-particle coordinates in a given direction as defined by 
Eq. \eqref{vec}, and $\vec{\bfi r}$ is a three dimensional vector formed by a set of single-particle coordinates.

A simple form of DECG functions are defined as
\begin{eqnarray}
   &&\exp \left\{
-\frac{1}{2} \tilde{{\bfi x}}A_{xx}^k{\bfi x}
-\frac{1}{2} \tilde{{\bfi y}}A_{yy}^k{\bfi y}
-\frac{1}{2} \tilde{{\bfi z}}A_{zz}^k{\bfi z}\right\} \nonumber\\
&\times&
\exp \left\{
-\tilde{{\bfi x}}A_{xy}^k{\bfi y}
-\tilde{{\bfi x}}A_{xz}^k{\bfi z}
-\tilde{{\bfi y}}A_{yz}^k{\bfi z}\right\} ,
\end{eqnarray}
where $A_{\alpha\beta}$ are $N\times N$ symmetric matrices. The scalar (inner) 
product $(\tilde {\bfi a}\cdot \tilde{\bfi b})$ for $N$-dimensional 
vectors $\tilde{\bfi a}=(a_1, a_2,...,a_N)$ and $\tilde{\bfi b}
=(b_1, b_2,..., b_N)$ is to be understood as  
$(\tilde {\bfi a}\cdot \tilde{\bfi b})=\sum_{m=1}^N a_mb_m$.  Assuming $A_{xx}=A_{yy}=A_{zz}=A$
and $A_{xy}=A_{xz}=A_{yz}=0$, one gets back the original definition of ECGs. 

Now we can define the block matrix $A$ as
\begin{equation}
    A=\left(
\begin{array}{ccc}
A_{xx}&A_{xy}&A_{xz}\\
A_{xy}&A_{yy}&A_{yz}\\
A_{xz}&A_{yz}&A_{zz}
\end{array}
\right),
\label{block}
\end{equation}
and the DECG function can be written as
\begin{equation}
\exp \left\{
-\frac{1}{2} \vec{{\bfi r}}A^k\vec{{\bfi r}}\right\},
\end{equation}
where the tilde is dropped for simplicity. The superscript $k$ stands for the $k$-th
basis function and
\begin{equation}
    \vec{{\bfi r}}A^k\vec{{\bfi r}}=\sum_{i,j=1}^{3N} r_i A^k_{ij}r_j.
\end{equation}

We multiply the simple DECG by 
\begin{equation}
\exp \left\{\vec{{\bfi r}}\vec{{\bfi s}}\right\}=
\exp \left\{    \sum_{i=1}^{3N} s_i r_i\right\},
\end{equation}
to form a basis that can describe nonzero angular momentum states and 
systems of multiple centers (molecules):
\begin{equation}
\label{psik}
\Psi_k=\exp\left\{
-\frac{1}{2} \vec{{\bfi r}}A^k\vec{{\bfi r}}+\vec{{\bfi r}}\vec{{\bfi s}}^k\right\}.
\end{equation}

As an example, assume that we write the trial function in the following form
\begin{widetext}
\begin{eqnarray}
\label{sums}
\label{decgform}
    &&{\rm exp}\left\{
-{\frac{1}{2}}\sum_{i<j}^N \alpha_{ij}^{xx}(x_i-x_j)^2
-{\frac{1}{2}}\sum_{i<j}^N \alpha_{ij}^{yy}(y_i-y_j)^2
-{\frac{1}{2}}\sum_{i<j}^N \alpha_{ij}^{zz}(z_i-z_j)^2
\right\} \\
&&\times{\rm exp}\left\{
-{\frac{1}{2}}\sum_{i,j=1}^N \alpha_{ij}^{xy}(x_i-y_j)^2
-{\frac{1}{2}}\sum_{i,j=1}^N \alpha_{ij}^{xz}(x_i-z_j)^2
-{\frac{1}{2}}\sum_{i,j=1}^N \alpha_{ij}^{yz}(y_i-z_j)^2
-{\frac{1}{2}}\sum_{i=1}^N \beta_i({\vec{r}_i-\vec{c}_i})^2
\right\}.
\nonumber
\end{eqnarray}
\end{widetext}
In this case, we have a correlation between the particle coordinates and
a single particle function centered at $\vec{c}_i$. The relation between the coefficients in Eq. \eqref{block} and Eq. \eqref{decgform} is shown in Appendix \ref{transform}.

\subsection{Hamiltonian} 
\label{A}
The Hamiltonian of the system is
\begin{equation}
    H=H_e+H_{ph}=H_e+H_p+H_{ep}+H_d.
\end{equation}
$H_e$ is the electronic Hamiltonian, and
$H_p$ is the photon Hamiltonian. The electron-photon coupling is
denoted as $H_{ep}$, and the dipole self-interaction is $H_d$.
In this case the electron-photon interaction is described by using the PF nonrelativistic QED Hamiltonian. 
The PF Hamiltonian can be derived
\cite{Ruggenthaler2018,Rokaj_2018,Mandal,acs.jpcb.0c03227,PhysRevB.98.235123} by
applying the Power-Zienau-Woolley gauge transformation \cite{Zienau}, 
with a unitary phase transformation on the minimal coupling ($p\cdot A$) Hamiltonian in the Coulomb gauge, 
\begin{equation}
    H_{ph}={\frac{1}{2}} \sum_{\alpha=1}^{N_p} \left[
p_{\alpha}^2+\omega_\alpha^2\left(
q_{\alpha}-\frac{{\vec{\lambda}_{\alpha}}} {{\omega_\alpha}}\cdot\vec{D}\right)^2
\right],
\end{equation}
where $\vec{D}$ is the dipole operator. The photon fields are described by quantized oscillators. 
$q_\alpha={\frac{1}{ \sqrt{2\omega_\alpha}}}(\hat{a}^+_\alpha+\hat{a}_\alpha)$ is the displacement field and 
$p_\alpha$ is the conjugate momentum.
This Hamiltonian describes $N_p$ photon modes with frequency
$\omega_{\alpha}$ and coupling  $\vec{\lambda}_{\alpha}$. 
The coupling term is usually written as \cite{PhysRevA.90.012508}
\begin{equation}
    \vec{\lambda}_{\alpha}=\sqrt{4\pi}\,S_\alpha(\vec{r})\vec{e}_\alpha,
\end{equation}
where $S_\alpha(\vec{r})$ is the mode function at position $\vec{r}$
and $\vec{e}_\alpha$ 
is the transversal polarization vector of the photon modes.

The electronic Hamiltonian is the usual Coulomb Hamiltonian and the  three components of the electron-photon 
interaction are as follows: The photonic part is
\begin{equation}
H_{p}=\sum_{\alpha=1}^{N_p}\left(\frac{1}{2} p_{\alpha}^{2}+\frac{\omega_{\alpha}^{2}}{2} q_{\alpha}^{2}\right) = 
\sum_{\alpha=1}^{N_p} \omega_{\alpha}\left(\hat{a}_{\alpha}^{+} \hat{a}_{\alpha}+\frac{1}{2}\right),
\label{phh}
\end{equation}
and the interaction term is
\begin{equation}
    H_{ep}=-\sum_{\alpha=1}^{N_p}\omega_{\alpha}q_\alpha
    \vec{\lambda}_{\alpha}\cdot\vec{D}=
    -\sum_{\alpha=1}^{N_p}\sqrt{\frac{\omega_{\alpha}}{
    2}}(\hat{a}_{\alpha}+\hat{a}_{\alpha}^+)\vec{\lambda}_{\alpha}\cdot\vec{D}.
\label{hep}
\end{equation}
Only photon states $|n_{\alpha}\rangle$, $|n_{\alpha}\pm 1\rangle$ are connected by $\hat{a}_{\alpha}$ 
and $\hat{a}_{\alpha}^+$. The matrix elements 
of the dipole operator $\vec{D}$ are only nonzero between spatial basis functions 
with angular momentum $l$ and $l\pm 1$ in 3D or $m$ and $m\pm 1$ in 2D.

The dipole self-interaction is defined as
\begin{equation}
H_{d}={\frac{1}{2}} \sum_{\alpha=1}^{N_p} \left(\vec{\lambda_{\alpha}}
\cdot \vec{D}\right)^{2},
\label{dsh}
\end{equation}
and the  importance of this term for the existence of a ground state is discussed in Ref. \cite{Rokaj_2018}.

In the following, we will assume that there is only one important
photon mode with frequency $\omega$ and coupling $\vec{\lambda}$. Thus the suffix $\alpha$ is omitted in what follows. The
formalism can be easily extended for many photon modes but here we
concentrate on calculating the matrix elements and it is sufficient to
use a single-mode. 

For one photon mode Eqs. \eqref{phh} \eqref{hep} and \eqref{dsh} can be simplified and the Hamiltonian becomes
\begin{equation}
    H=T+V+U+\omega\left(\hat{a}^+\hat{a}+{1\over 2}\right)+\omega\vec{\lambda}\cdot\vec{D}q+{\frac{1}{2} }(\vec{\lambda}\cdot\vec{D})^2,
    \label{Hamil}
\end{equation}
where $T$ is the kinetic operator
\begin{equation}
    T=-\frac{1}{2} \sum_{i=1}^{N}\left(\frac{\partial^{2}}{\partial x_{i}^{2}}+\frac{\partial^{2}}{\partial y_{i}^{2}}+\frac{\partial^{2}}{\partial z_{i}^{2}}\right).
    \label{kin}
\end{equation}
$V$ is the Coulomb interaction
\begin{equation}
    V=\sum_{i<j} V_c(\vec{r}_i-\vec{r}_j),  \ \ \ \ \ \ V_c(\vec{r}_i-\vec{r}_j)={\frac{q_iq_j}{|\vec{r}_i-\vec{r}_j|}}.
\end{equation}
$U$ is an external potential 
\begin{equation}
    U=\sum_{i=1}^N U(\vec{r}_i),
\end{equation}
and the dipole moment $\vec{D}$ of the system is defined as
\begin{equation}
    \vec{D}=\sum_{i=1}^N q_i\vec{r}_i.
\end{equation}

The operators act in real space, except $q$ which acts on the photon space
\begin{eqnarray}
q|n\rangle&=&\frac{1}{\sqrt{2
\omega}}\left(a+a^{+}\right)|n\rangle\\
&=&
\frac{1}{\sqrt{2 \omega}}\left(\sqrt{n}|
n-1\rangle+\sqrt{n+1}|n+1\rangle\right)\nonumber .
\end{eqnarray}

In the following, we calculate the matrix elements of DECGs. 
Most of the matrix elements calculated previously for  ECGs remain 
the same except the one that uses the DECG block matrix. 

\subsection{Overlap matrix}

The overlap matrix is given by 
\begin{equation}
\langle\Psi_i\vert\Psi_j\rangle=\int
\exp \left\{-\frac{1}{2} {\vec{\bfi r}}A{\vec{\bfi r}}+{\vec{\bfi r}}{\vec{\bfi s}}\right\}
d{\vec{\bfi r}},
\end{equation}
where $A$ and $\bfi{\vec{s}}$ are defined as
\begin{equation}
    A=A^i+A^j \ \ \ \ \ \vec{{\bfi s}}=\vec{{\bfi s}}^{\,i}+\vec{{\bfi s}}^{\,j},
\end{equation}
and can be calculated using Eq. \eqref{ggaussint1} in Appendix \ref{ecg}: 
\begin{equation}
    \langle\Psi_i\vert\Psi_j\rangle={\frac{(2\pi)^{3N/2}}{ ({\rm det} A)^{1/2}}}
    \exp \left\{\frac{1}{2} \vec{{\bfi s}}A^{-1}\vec{{\bfi s}}\right\}.
    \label{overlap}
\end{equation}

\subsection{Kinetic energy}
We will write the kinetic energy operator in the following form
\begin{equation}
    T={\vec{\bfi p}}\Lambda{\vec{\bfi p}},
\end{equation}
where the momentum operator is given by
\begin{equation}
    p_i=-i\hbar {\frac{\partial}  {\partial r_i}} \ \ \ \ \ (i=1,...,3N).
\end{equation}
For a system of particles with masses $m_1,...,m_N$, $\Lambda$ is
a block diagonal matrix 
\begin{equation}
    \Lambda=\left(
\begin{array}{ccc}
\Lambda^x&0        &0\\
0        &\Lambda^y &0\\
0        &  0       &\Lambda^z
\end{array}
\right),
\end{equation}
where the matrix elements of the block diagonal matrix are given by 
\begin{equation}
\Lambda^{\alpha}_{ij}={\frac{1}{2 m_i}} \delta_{ij},
\end{equation}
for systems where the external potential fixes the center of the system 
(e.g.  electrons in a harmonic oscillator potential, or electrons in an atom 
where the mass of the nucleus is taken to be infinity). Otherwise, we have to remove the center of mass motion  of the system using 
\begin{equation}
\Lambda^{\alpha}_{ij}={\frac{1}{2 m_i}} \delta_{ij}-{\frac{1}{2M}}, 
\end{equation}
where $M=m_1+m_2+...m_N$. In principle $\Lambda^x$, $\Lambda^y$ and $\Lambda^z$
can be different if the masses of particles depend on the directions.

Taking the derivative on the right-hand side
\begin{widetext}

\begin{equation}
    {\frac{\partial}  {\partial r_i}} \exp \left\{-\frac{1}{2} \vec{{\bfi r}}A^j\vec{{\bfi r}} + {\vec{\bfi r}}{\vec{\bfi s}}\right\}=
\left(-(A^j\vec {\bfi r})_i+s_i\right)
\exp \left\{-\frac{1}{2} \vec{{\bfi r}}A^j\vec{{\bfi r}} + {\vec{\bfi r}}{\vec{\bfi s}} \right\}.
\end{equation}

Using analogous results on the left side, the overlap with the kinetic energy operator can be given by 

\begin{eqnarray}
\langle\Psi_i\vert T\vert\Psi_j\rangle&=&\int
\left(\vec{{\bfi r}} (A^i\Lambda A^j)
\vec{{\bfi r}}+
\vec{{\bfi s}}^{\,i} \Lambda \vec{{\bfi s}}^{\,j}-
\vec{{\bfi s}}^{\,i} \Lambda A^j\vec{{\bfi r}}-
A^i\vec{{\bfi r}} \Lambda \vec{{\bfi s}}^{\,j}
\right)
\exp \left\{-\frac{1}{2} \vec{{\bfi r}}A\vec{{\bfi r}} + {\vec{\bfi r}}{\vec{\bfi s}} \right\}
d\vec{\bfi r}\nonumber \\
&=& 
\left({\rm Tr}(A^i\Lambda A^kA^{-1})-
\vec{{\bfi y}} \Lambda \vec{{\bfi y}}
\right)
\langle\Psi_i\vert\Psi_j\rangle,
\end{eqnarray}
\end{widetext}
where we used Eqs. \eqref{ggaussint2} and \eqref{ggaussint3} and define $\bfi{\vec{y}}$ as
\begin{equation}
    \vec{{\bfi y}}=A^jA^{-1}{\vec{\bfi s}}^{\,i}-A^iA^{-1}{\vec{\bfi s}}^{\,j}.
\end{equation}

\subsection{Potential energy}
Both $V_c$ and $U$ can be rewritten using a $\delta$ function,
\begin{equation}
    V_c(\vec{r}_i-\vec{r}_j)=\int \delta(\tilde{w}^{ij}\vec{\bfi r}-{\vec r}) V_c(\vec{r})d\vec{r},
\end{equation}
where $\tilde{w}^{ij}\vec{\bfi r}$ is a short-hand notation for 
$\sum_{k=1}^N w^{ij}_k{\vec r}_k$ and in this case $w^{ij}_k=\delta_{ik}-\delta_{jk}$. The corresponding formula for $U$ is
\begin{equation}
    U(\vec{r}_i)=\int \delta(\tilde{w}^{i}\vec{\bfi r}-{\vec r}) U(\vec{r})d\vec{r},
    \label{externalpot}
\end{equation}
with $w^i_k=\delta_{ik}$. This form allows us to calculate the matrix elements
for $\delta(\tilde{w}\vec{\bfi r}-{\vec r})$ for a general case without using the
particular form of the potential, and to calculate the matrix element of the 
potential by integration over ${\vec r}$. The $\delta$ function can be represented by
(we drop the superscript $ij$ and $i$ of $w$ for simplicity)
\begin{equation}
    \delta(\tilde{w}\vec{\bfi r}-{\vec r})={\frac{1}{ (2\pi)^3}}\int {\rm e}^{i\vec{k}
    (\tilde{w}\vec{\bfi r}-{\vec r})} d{\vec{k}}.
\end{equation}
We want to calculate the matrix elements
\begin{widetext}
\begin{equation}
\langle\Psi_i\vert \delta(\tilde{w}\vec{\bfi r}-{\vec r})\vert\Psi_j\rangle=
{\frac{1}{(2\pi)^3}}
\int\int
    {\rm e}^{i\vec{k}
    (\tilde{w}\vec{\bfi r}-{\vec r})}
    \exp \left\{
-\frac{1}{2} \vec{{\bfi r}}A\vec{{\bfi r}} +\vec{{\bfi r}} \vec{{\bfi s}} \right\}
d\vec{\bfi r}d{\vec {k}}.
\end{equation}
\end{widetext}
This can be done by defining $\bfi{\vec{t}}$ as
\begin{equation}
   \vec{\bfi t}=
   \left(
\begin{array}{c}
ik_1\tilde{w}\\
ik_2\tilde{w}\\
ik_3\tilde{w}
\end{array}
\right)+\vec{{\bfi s}}.
\end{equation}
Using Eq. \eqref{ggaussint1}, we can express the matrix element as
\begin{widetext}
\begin{eqnarray}
\langle\Psi_i\vert \delta(\tilde{w}\vec{\bfi r}-{\vec r})\vert\Psi_j\rangle&=&
{\frac{1} {(2\pi)^3}}\left({\frac{(2\pi)^{3N}} { {\rm det} A }}\right)^{\frac{1}{2}}
\int {\rm e}^{-i\vec{k}{\vec r}}
{\rm exp}\Big({\frac{1}{2}}{{\vec{\bfi t}}} A^{-1} \vec{\bfi t}\Big)
  d{\vec {k}}\nonumber \\ &=&
 {\frac{1} {(2\pi)^3}}\left({\frac{(2\pi)^{3N}} { {\rm det} A }}\right)^{\frac{1}{2}}
\int {\rm e}^{-i\vec{k}{\vec r}}
{\rm exp}\Big(-{\frac{1}{2}}{{\vec{k}}} B \vec{ k}
+{\frac{1}{2}}{{\vec{\bfi s}}} A^{-1} \vec{\bfi s}
+i{{\vec{k}}} \vec{ b}
\Big)
  d{\vec k},
\end{eqnarray}
\end{widetext}
where $B$ is a $3\times 3$ matrix given by
\begin{equation}
    B=\left(
\begin{array}{ccc}
B_{11}&B_{12}&B_{13}\\
B_{12}&B_{22}&B_{23}\\
B_{13}&B_{23}&B_{33}
\end{array}
\right),
\end{equation}
with the matrix elements of $B$ defined as
\begin{equation}
    B_{ij}=\sum_{k=(i-1)\cdot N+1}^{i\cdot N} \ \  \sum_{l=(j-1)\cdot N+1}^{j\cdot N}w_{k'}A^{-1}_{kl}w_{l'},
\end{equation}
where $k'=k-(i-1)\cdot N$ and $l'=l-(j-1)\cdot N$. We have also
defined a three dimensional vector ${\vec b}$:
\begin{equation}
    b_i=\sum_{k=(i-1)\cdot N+1}^{i\cdot N} w_{k'}\left(A^{-1}{\vec{\bfi s}}\right)_k.
\end{equation}

The last integral can again be calculated using Eq. \eqref{ggaussint1} and we have
\begin{widetext}
\begin{equation}
\langle\Psi_i\vert \delta(\tilde{w}\vec{\bfi r}-{\vec r})\vert\Psi_j\rangle=
{\frac{1} {(2\pi)^{3/2} ({\rm det} B)^{1/2} }} 
{\rm exp}\Big(-{\frac{1}{2}}(\vec {r}-{\vec {b}}) B^{-1} (\vec {r}-\vec{b})\Big)
\langle\Psi_i\vert\Psi_j\rangle. \ \ \ \ 
\label{potme}
\end{equation}
\end{widetext}
Integrating over $\vec r$ should give back the overlap, and using Eq. \eqref{ggaussint1} one immediately gets these results. 
Note that Eq. \eqref{potme} can also be used to calculate the 
single particle density. This formula is generalized for two particle density in Appendix \ref{2prob}.

\subsection{Electron-photon coupling}
\label{F}
By introducing $\bfi{\vec{q}}$ as
\begin{equation}
    \vec{\bfi q}=
\left(
\begin{array}{c}
\lambda_1 q_1\\
\lambda_1 q_2\\
\vdots \\
\lambda_1 q_N\\
\lambda_2 q_1\\
\vdots \\
\lambda_2 q_{N}\\
\lambda_3 q_1\\
\vdots\\
\lambda_3 q_{N}
\end{array}
\right),
\end{equation}
the relevant part of the  coupling term can be written as
\begin{equation}
    \vec{\lambda}\cdot \vec{D}={\vec{\bfi q}}{\vec{\bfi r}},
\end{equation}
and the matrix elements of this term can be easily calculated using
Eq. \eqref{ggaussint2}
\begin{eqnarray}
\langle\Psi_i\vert \vec{\lambda}\cdot \vec{D}\vert\Psi_j\rangle&=&\int
\vec{\lambda}\cdot \vec{D}
\exp \left\{-\frac{1}{2} \vec{{\bfi r}}A\vec{{\bfi r}}+{\vec{\bfi r}}{\vec{\bfi s}}\right\}
d\vec{\bfi r}\nonumber \\
&=&  \vec{{\bfi q}}A^{-1}\vec{{\bfi s}} \  \langle\Psi_i\vert\Psi_j\rangle.
\end{eqnarray}

\subsection{Dipole self-interaction}
The dipole self-interaction can also be readily available using Eq.
\eqref{ggaussint3}:
\begin{eqnarray}
\begin{aligned}
\langle\Psi_i\vert {\frac{1}{2}} (\vec{\lambda}\cdot \vec{D})^2\vert\Psi_j\rangle&=\int
{\frac{1}{2}}(\vec{\lambda}\cdot \vec{D})^2
\exp \left\{-\frac{1}{2} \vec{{\bfi r}}A\vec{{\bfi r}}+ 
{\vec{\bfi r}}{\vec{\bfi s}}\right\}
d\vec{\bfi r}\\
&={\frac{1}{2}} \left(
\vec{{\bfi q}}A^{-1}\vec{{\bfi q}}+
\left(\vec{{\bfi q}}A^{-1}\vec{{\bfi s}}\right)^2
\right)\langle\Psi_i\vert\Psi_j\rangle.
\end{aligned}
\end{eqnarray}

\subsection{Eliminating the dipole self-interaction}
One motivation of DECG is that the dipole self-interaction
term of the Hamiltonian can be eliminated using a special choice of
DECG exponentials producing a much simpler Hamiltonian.

The dipole self-interaction term is a special quadratic form and this 
quadratic form can be represented with DECG exponent. One can try to 
find a suitable $\alpha$ to eliminate the dipole self-interaction using
the kinetic energy operator:
\begin{equation}
-\frac{1}{2} \sum_{i=1}^{3N}\left(\frac{\partial^{2}}{\partial r_{i}^{2}}\right) \exp \left(\alpha(\vec{\lambda} \cdot \vec{D})^{2}\right).
\end{equation}
To solve this we need to evaluate the second derivative of the exponential with respect to $\vec{r}_i$. The first derivative with respect to $x_i$ is given by  
\begin{equation}
\frac{\partial}{\partial x_{i}} \exp \left(\alpha(\vec{\lambda} \cdot \vec{D})^{2}\right)=2 \alpha \lambda_{1}{q}_i(\vec{\lambda} \cdot \vec{D}) \exp \left(\alpha(\vec{\lambda} \cdot \vec{D})^{2}\right),
\end{equation}
and the second derivative
\begin{eqnarray}
\frac{\partial^{2}}{\partial x_{i}^{2}} \exp \left(\alpha(\vec{\lambda} \cdot \vec{D})^{2}\right)
&=&2 \alpha \lambda_{1}^{2} {q}_{i}^{2} \exp \left(\alpha(\vec{\lambda} \cdot \vec{D})^{2}\right)\\
&+&4 \alpha^{2} \lambda_{1}^{2} {q}_{i}^{2} (\vec{\lambda} \cdot \vec{D})^{2} \exp \left(\alpha(\vec{\lambda} \cdot \vec{D})^{2}\right),
\nonumber
\end{eqnarray}
with similar expressions for $y_i$ and $z_i$.
By choosing $\alpha$ as
\begin{equation}
\alpha=\frac{1}{2 \sqrt{\sum_{i=1}^{N}{q}_{i}^{2}}\,\lambda},
\label{alpha}
\end{equation}
where $\lambda$ is the magnitude of $\vec{\lambda}$,
we can express the kinetic energy operator acting on the exponential as
\begin{widetext}
\begin{equation}
\label{self}
-\frac{1}{2} \sum_{i=1}^{3N}\left(\frac{\partial^{2}}{\partial r_{i}^{2}}\right) \exp \left(\alpha(\vec{\lambda} \cdot \vec{D})^{2}\right)=
-\left({\frac{1} {4\alpha}}+{\frac{1}{2}}(\vec{\lambda} \cdot \vec{D})^{2}\right) \exp \left(\alpha(\vec{\lambda} \cdot \vec{D})^{2}\right).
\end{equation}
\end{widetext}
This means that by multiplying the basis with the factor
\begin{equation}
\exp \left(\alpha(\vec{\lambda} \cdot \vec{D})^{2}\right),
\end{equation}
the dipole self-interaction can be removed and the numerical solution is
much simpler. In this way, the nonspherical dipole self-interaction
is eliminated. In other words, it is built in the basis functions.
Note that the above exponential form can be recast into a DECG, 
but not into an ECG. The generalization of Eq. \ref{self} to multiphoton mode can be found in Appendix \ref{gendip}.

\section{Numerical Examples}
\label{III}
In this section, we present a few numerical examples to show that the matrix
elements evaluated in this paper can be used in practical calculations. 
We will not fully explore the efficiency of the DECG basis, and we restrict our 
approach to an $A$ matrix of the form
\begin{equation}
    A=\left(
\begin{array}{ccc}
A_{xx}&  0   &0\\
0     &A_{xx}&0\\
0     & 0    & A_{xx}
\end{array}
\right),
\end{equation}
and the trial function is
\begin{equation}
\label{psik2}
\Psi_k=\exp\left\{
-\frac{1}{2} \vec{{\bfi r}}(A^k+2\alpha(\vec{\lambda}\cdot\vec{D}))\vec{{\bfi r}}+\vec{{\bfi r}}\vec{{\bfi s}}^k\right\},
\end{equation}
where $\alpha$ is defined in Eq. \eqref{alpha}. If $\alpha=0$ then this 
function is the conventional ECG basis function. Nonzero $\alpha$ leads to
nonzero off diagonal block matrices and the basis becomes DECG. 

In these calculations, we have used the separable approximation of $1/r$ in terms of Gaussians \cite{BEYLKIN200517}
\begin{equation}
    {\frac{1}{r}}=\sum_k w_k{\rm e}^{-p_k r^2}.
    \label{gauss-hermite}
\end{equation}
In this way, the integral in Eq. \eqref{potme} can be analytically evaluated
(a numerical approach is presented in Appendix \ref{potmatrix}). 
89 Gaussian functions with the coefficients $w_k$ and $p_k$ (taken from
Ref. \cite{BEYLKIN200517}) can approximate $1/r$ with an error less than
$10^{-8}$ in the interval $[10^{-9},1]$a.u. For larger intervals, one can easily 
scale to coefficients. Note that this expansion uses significantly
fewer terms than a Gaussian quadrature for the same accuracy \cite{BEYLKIN200517}.

\begin{figure}
\includegraphics[width=0.45\textwidth]{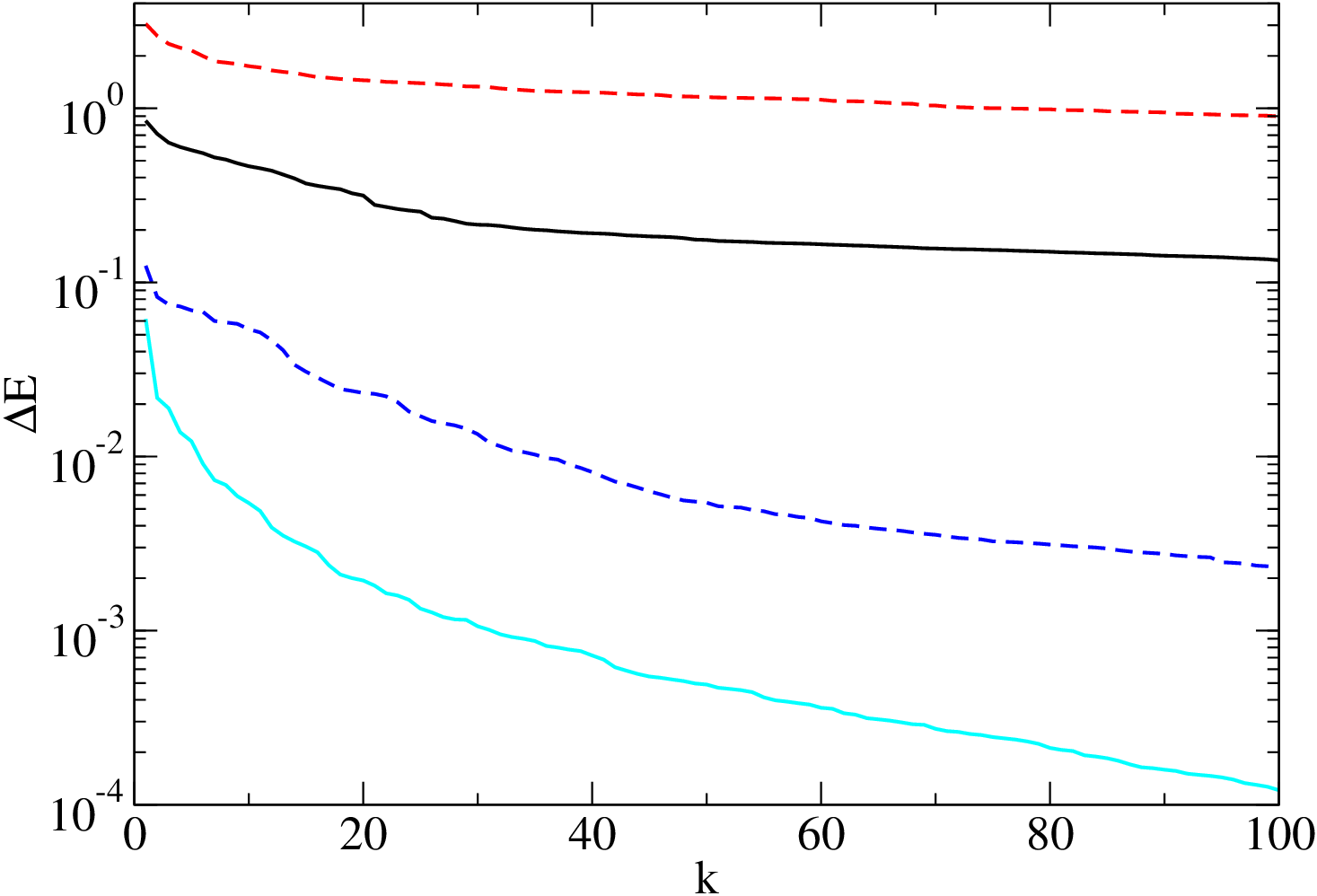}
\caption{Energy convergence as a function of basis dimension. $\Delta E$ is the difference of the calculated energy and  the exact energy. The lower two curves are DECG calculations for $\lambda=1$ a.u. (solid line), $\lambda=2.5$ a.u. (dashed line); the upper two curves are ECG calculations for  $\lambda=1$ a.u. (solid line), $\lambda=2.5$ a.u. (dashed line). }
\label{2e}
\end{figure}

As a first example, we consider a 2D system of 2 electrons in a harmonic oscillator confinement potential,
\begin{equation}
{\frac{1}{2}} \omega_0^2 \sum_{i=1}^2 \vec{r}_i^{\,2},
\end{equation}
interacting via a Coulomb potential. This problem is analytically solvable
\cite{henry} and we will compare the ECG ($\alpha=0$) and DECG solution. We take
$\omega=0$ in Eq. \eqref{Hamil}, so there is no coupling to photons but the
potential is nonspherical because $\lambda\ne0$. We test two $\lambda$ values:
$\lambda=1$ a.u. (the energy is $E=2.7807764$ a.u.) and $\lambda=2.5$ ($E=4.2624689$). Fig. \ref{2e} shows the convergence of energy as a function 
of the number of basis states. Each basis state is selected by comparing 250 random parameter sets and choosing the one that minimizes the energy. The DECG
converges up to 3-4 digits on a basis of 100 states. The ECG converges much slower, and for the stronger coupling ($\lambda=2.5$) the energy is 0.9 a.u. above the exact value. A larger basis dimension and more parameter optimizations
would improve the results, but this already shows the general tendency and the superiority of the DECG basis. Note, that the 
ECG would also converge to the exact result after more optimization and much larger basis size. 

The next example is a 2D H$_2$ molecule with nuclei fixed  at distance $r$. In this case, we assume that there is only one relevant photon mode with frequency $\omega=1.5$ a.u. There are infinitely many photons with energy $n\hbar\omega$ ($n=0,1,2,...)$, but only the lowest photon  states are coupled to the electronic part. We solver Eq. \eqref{Hamil}  
using the lowest $n=0,..,5$ photon spaces. The energy of a 2D H atom is $E=-2$ a.u. without coupling the photons. The energy of the atom coupled to photons with $\lambda=$1.5 a.u. increases to $E=-1.71$ a.u. The increase is largely due to the dipole self energy part
in Eq. \eqref{Hamil}. The probability amplitudes of the spatial wave function in photon spaces are 0.988 ($n=0$), 0.01 ($n=1)$ and 0.001 $n=2$.  These are small probabilities but there is a relatively strong coupling between the electrons and light. This is shown by the fact that the energy without coupling (solely due to the dipole self-interaction and the Coulomb) is -1.67 a.u. By increasing the 
coupling further the energy of the H$_2$ increases (e.g. for $\lambda=3$, $E$=-1.15 a.u.). 

\begin{figure}
\includegraphics[width=0.45\textwidth]{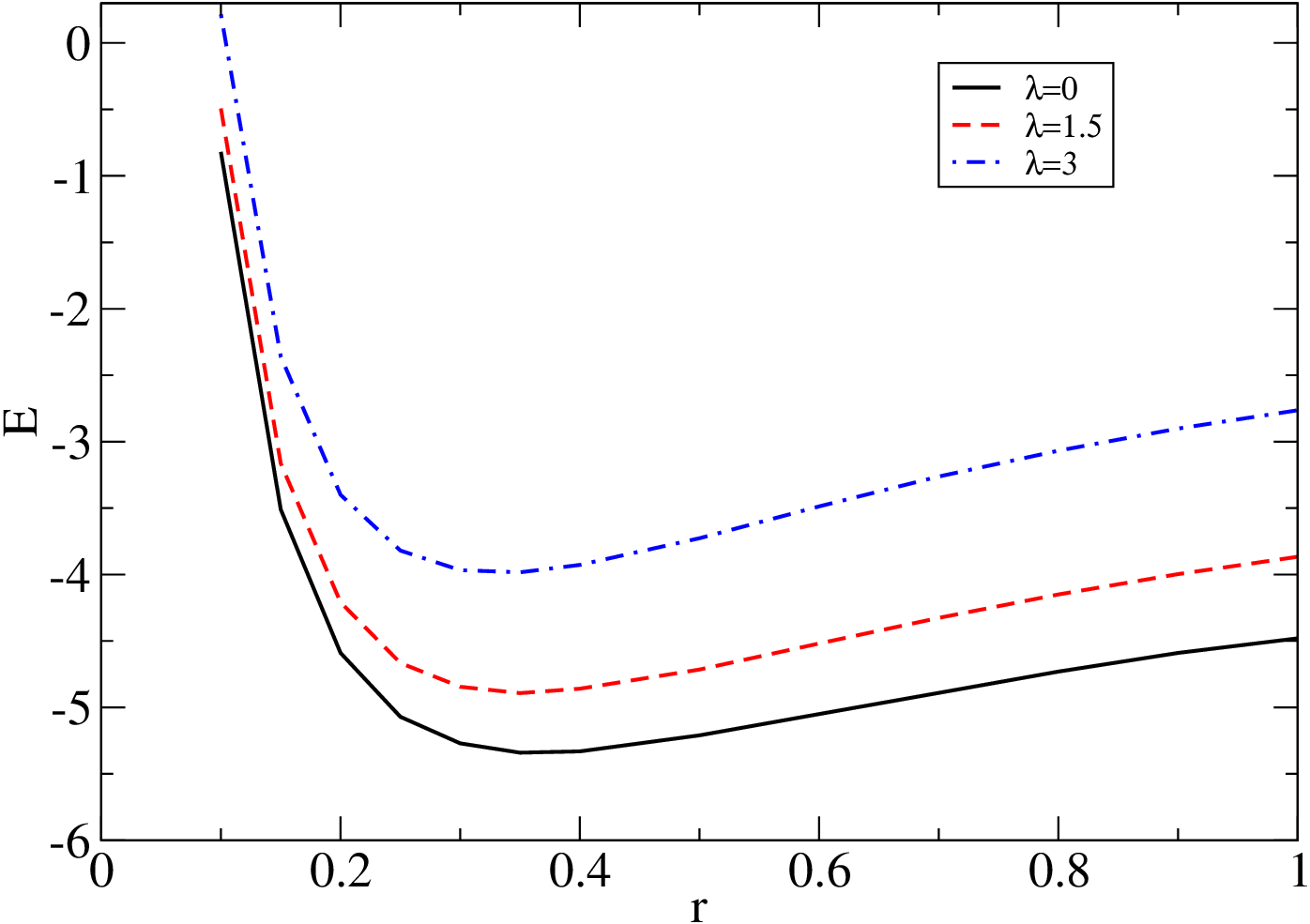}
\caption{Energy of the 2D H$_2$ molecule as a function of the proton-proton distance.}
\label{h2}
\end{figure}

Fig. \ref{h2} shows the  energy of the H$_2$ molecule with and without coupling to light. Without coupling to light, the 2D H$_2$ molecule has a
lowest energy at around $r$=0.35 a.u. The the H$_2$ molecule to light the energy minimum slightly shifts toward shorter distances. Overall the three curves are very similar except that the dipole self-interaction term pushes them higher with increasing $\lambda$. The binding energies at the minimum energy point increase with $\lambda$: $E_b$=1.34 a.u. ($\lambda=0$),
$E_b=1.47$ ($\lambda=1.5$ a.u.), and $E_b=1.68$ a.u. ($\lambda=3$ a.u.), where $E_b$ is the difference of the energy of the molecule and two times the energy
of the H atom. 

\begin{figure}
\includegraphics[width=0.45\textwidth]{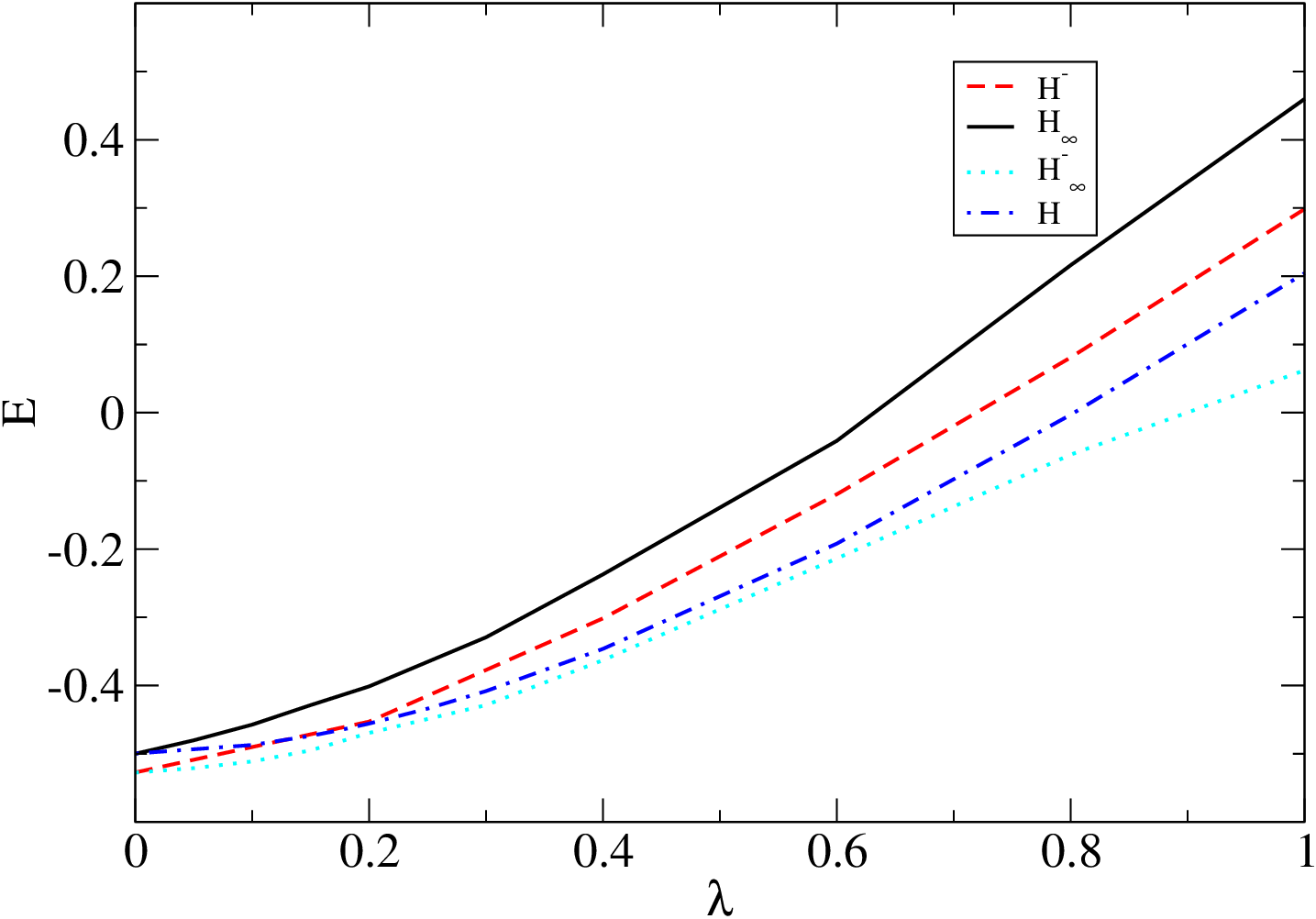}
\caption{Energy of the 3D H$^-$ ion as a function of $\lambda$ ($\omega=0.8$ a.u.}
\label{hm}
\end{figure}

The final example is the H$^-$ ion with finite ($m_H=1836.1515$ a.u.) and infinite nuclear mass in 3D. Fig. \ref{hm} shows the energy of the H atom and H$^-$ ion as a function of $\lambda$. As the figure shows, the H$^-$ dissociates for strong $\lambda$ in the finite mass case but remains stable in the infinite mass case. In the finite mass case, the dissociation happens around $\lambda=0.08$ a.u., at that point, the energy of the H plus an electron system becomes lower than that on H$^-$ (the energy of the electron
coupled to light is calculated by solving Eq. \eqref{Hamil} for the electron).
This example shows the importance of explicit treatment of the system
as a three-body system because the light strongly couples to the proton as well.

\section{Summary}
We have introduced a new variant of ECG basis functions that are suitable
for problems with nonspherical potentials. All necessary matrix elements 
are calculated and numerically tested. The treatment of the Coulomb interaction is more complicated than in the conventional ECG case due to the
nonspherical integrals that appear in the interaction part. Two ways are proposed to solve this problem. One can either expand the Coulomb potential
in Gaussians and the integration becomes analytical, or use numerical integration. 

We have shown that using the DECG basis the coupled light-matter equations can be efficiently solved even  when the coupling and thus the dipole self-interaction term is large. This opens up the way to calculate light-matter coupled few-body systems with high accuracy in cavity QED systems. 

The approach might be useful in other problems with nonspherical potentials e.g. calculation of atoms and molecules in magnetic fields.

\appendix
\section{Relating different forms of basis functions}
\label{transform}
In this appendix we show how Eq. \eqref{psik} and Eq. \eqref{decgform} 
can be related and how the matrix elements of the trial function can be determined.
\begin{widetext}
\begin{equation}
\begin{array}{l}
\begin{aligned}
\Psi&=\exp \left\{-\frac{1}{2} \sum_{i<j}^{N} \alpha_{i j}^{x x}\left(x_{i}-x_{j}\right)^{2}-\frac{1}{2} \sum_{i<j}^{N} \alpha_{i j}^{y y}\left(y_{i}-y_{j}\right)^{2}-\frac{1}{2} \sum_{i<j}^{N} \alpha_{i j}^{z z}\left(z_{i}-z_{j}\right)^{2}\right\} \\
&\times \exp \left\{-\frac{1}{2} \sum_{i,j=1}^{N} \alpha_{i j}^{x y}\left(x_{i}-y_{j}\right)^{2}-\frac{1}{2} \sum_{i,j=1}^{N} \alpha_{i j}^{x z}\left(x_{i}-z_{j}\right)^{2}-\frac{1}{2} \sum_{i,j=1}^{N} \alpha_{i j}^{y z}\left(y_{i}-z_{j}\right)^{2}-\frac{1}{2} \sum_{i=1}^{N} \beta_{i}\left(\vec{r}_{i}-\vec{c}_{i}\right)^{2}\right\}.
\end{aligned}
\end{array}
\end{equation}
\end{widetext}
The diagonal blocks of the trial function evaluate to
\begin{equation}
\sum_{i<j}^{N} \alpha_{i j}^{x x}\left(x_{i}-x_{j}\right)^{2}={\bfi{x}} M_{x x} \bfi{x},
\end{equation}
where $ M_{x x} $ is an $ N \times N $ symmetric matrix:
\begin{equation}
\left(M_{x x}\right)_{i i}=\sum_{\substack{k=1 \\ k \neq i}}^{N} \alpha_{i k}^{x x}, \quad\left(M_{x x}\right)_{i j}=-\alpha_{i j}^{x x} \quad \text { for } i \neq j.
\end{equation}
Here, $ \alpha_{j i}^{x x} $ for $ j>i $ is set equal to $ \alpha_{i j}^{x x} $.
The off-diagonal blocks of the trial function evaluate to
\begin{equation}
\sum_{i,j=1}^{N} \alpha_{i j}^{x y}\left(x_{i}-y_{j}\right)^{2}={\bfi{x}} G_{x x}^{x y} \bfi{x}+{\bfi{y}} G_{y y}^{x y} \bfi{y}+{\bfi{x}} G_{x y}^{x y} \bfi{y}+{\bfi{y}} {G_{x y}^{x y}} \bfi{x},
\end{equation}
where $ G_{x x}^{x y} $ and $ G_{y y}^{x y} $ are both diagonal,
\begin{equation}
\begin{aligned}
\left(G_{x x}^{x y}\right)_{i i}&=\sum_{j=1}^{N} \alpha_{i j}^{x y},\\
\left(G_{y y}^{x y}\right)_{i i}&=\sum_{i=1}^{N} \alpha_{i j}^{x y},
\end{aligned}
\end{equation}
while $ G_{x y}^{x y} $ is defined as
\begin{equation}
\left(G_{x y}^{x y}\right)_{i j}=-\alpha_{i j}^{x y}(\text { for } i,j= 1, \ldots, N). \\ 
\end{equation}

The single-particle product element of the trial function is
\begin{equation}
\begin{aligned}
\sum_{i=1}^{N} \beta_{i}\left(\vec{r}_{i}-\vec{c}_{i}\right)^{2} &=\sum_{i=1}^{N} \beta_{i}\left[\left(x_{i}^{2}+y_{i}^{2}+z_{i}^{2}\right)-2 \vec{c}_{i} \cdot \vec{r}_{i}+\vec{c}_{i} \cdot \vec{c}_{i}\right]\\
&={\bfi{x}} B \bfi{x}+{\bfi{y}} B \bfi{y}+{\bfi{z}} B \bfi{z}\\ &-2 \sum_{i=1}^{N} \beta_{i} \vec{c_{i}} \cdot \vec{r_{i}}+\sum_{i=1}^{N} \beta_{i} \vec{c_{i}} \cdot \vec{c_{i}},
\end{aligned}
\end{equation}
where $ B $ is an $ N \times N $ diagonal matrix with $ B_{i i}=\beta_{i} $, and
\begin{equation}
\sum_{i=1}^{N} \beta_{i} \vec{c_{i}} \cdot \vec{r_{i}}=\sum_{i=1}^{N} \beta_{i}\left(c_{i_{x}} x_{i}+c_{i_{y}} y_{i}+c_{i_{z}} z_{i}\right)={\vec{\bfi s}} \cdot \vec{\bfi{r}}.
\end{equation}
\begin{widetext}
Here, ${\vec{\bfi s}}=\left(\beta_{1} c_{1_{x}}, \beta_{2} c_{2_{x}}, \ldots, \beta_{N} c_{N_x}, \beta_{1} c_{1_{y}}, \beta_{2} c_{2_{y}}, \ldots, \beta_{N} c_{N_y}, \beta_{1} c_{1_{z}}, \beta_{2} c_{2_{z}}, \ldots, \beta_{N} c_{N_z}\right) . $ Combining the above results
leads to
\begin{equation}
\Psi=\exp \left\{-\frac{1}{2} {\vec{\bfi{r}}} A \vec{\bfi{r}}+{\vec{\bfi s}} \vec{\bfi{r}}-\frac{1}{2} \sum_{i=1}^{N} \beta_{i} \vec{c_{i}} \cdot \vec{c_{i}}\right\},
\end{equation}
with the matrix $A$ given by
\begin{equation}
A=\left(\begin{array}{ccc}
M_{x x}+G_{x x}^{x y}+G_{x x}^{x z}+B & G_{x y}^{x y} & G_{x z}^{x z} \\
G_{x y}^{x y} & M_{y y}+G_{y y}^{x y}+G_{y y}^{y z}+B & G_{y z}^{y z} \\
G_{x z}^{x z} & G_{y z}^{y z} & M_{z z}+G_{z z}^{x z}+G_{z z}^{y z}+B
\end{array}\right).
\end{equation}
\end{widetext}
Comparing $ A $ with Eq. \eqref{block}, one obtains $ A_{\alpha \beta} $. For example,
\begin{equation}
A_{x x}=M_{x x}+G_{x x}^{x y}+G_{x x}^{x z}+B, \quad A_{x y}=G_{x y}^{x y},
\end{equation}
and their matrix elements are easily obtained from the above defining equations.

\section{Generalized Gaussian integrals}
\label{ecg}
In this appendix, we list the most important formulas for Gaussian integrals.
These results are valid in $n=d\times N $dimension, where $d$ is the space dimension.
Define the generating function

\begin{equation}
g(\vec{\bfi s}; A,\vec{\bfi r})=
{\rm exp}\Big(-{\frac{1}{2}}{{\vec{\bfi r}}}A\vec{\bfi r}+{\vec{\bfi s}}\vec{\bfi r}
\Big). 
\label{gfn}
\end{equation}

The evaluation of a Gaussian integral of this form is given by

\begin{equation}
\int  {\rm exp}\Big(-{\frac{1}{2}}{{\vec{\bfi r}}}A \vec{\bfi r}+
{{\vec{\bfi s}}}\vec{\bfi r}\Big)d\vec{\bfi r}
=\left({\frac{(2\pi)^{n}} { {\rm det} A }}\right)^{\frac{1}{2}}
{\rm exp}\Big({\frac{1}{2}}{{\vec{\bfi s}}} A^{-1} \vec{\bfi s}\Big), 
\label{ggaussint1}
\end{equation}

Some useful formulas related to this integral are collected below. 
By differentiating both sides of the above equation 
with respect to the $i$th component of 
the vector $\vec{\bfi s}$, $s_i$, we obtain 
\begin{eqnarray}
\label{ggaussint2}
& &\int  r_i \,{\rm exp}\Big(-{\frac{1}{2}}
{{\vec{\bfi r}}}A\vec{\bfi r}+{{\vec{\bfi s}}}\vec{\bfi r}\Big)d\vec{\bfi r}
\quad \\ \nonumber  & & =\, (A^{-1}\vec{\bfi s})_i \left({\frac{(2\pi)^{n}} { {\rm det} A }}
\right)^{\frac{1}{2}}
{\rm exp}\Big({\frac{1}{2}}{{\vec{\bfi s}}} A^{-1} \vec{\bfi s}\Big). 
\end{eqnarray}
Further differentiation with respect to $\vec{\bfi s}_j$ leads us to 
\begin{eqnarray}
\label{ggaussint3}
&&\int  r_i r_j {\rm exp}\Big(-{\frac{1}{2}}
{{\vec{\bfi r}}}A \vec{\bfi r}+{{\vec{\bfi s}}}\vec{\bfi
r}\Big)d\vec{\bfi r} =\left({\frac{(2\pi)^{n}} { {\rm det} A }}\right)^{\frac{1}{2}} \\
&\times&
{\rm exp}\Big({\frac{1}{2}}{{\vec{\bfi s}}} A^{-1} \vec{\bfi
s}\Big) \Big\{(A^{-1})_{ij}+(A^{-1}\vec{\bfi s})_i
(A^{-1}\vec{\bfi s})_j\Big\} \nonumber
\end{eqnarray}
\begin{widetext}
\begin{table}
\caption{Matrix elements, ${\cal M}=\langle 
g(\vec{\bfi s}'; A',\vec{\bfi r})| {\cal O}| g(\vec{\bfi s}; A,\vec{\bfi r})\rangle$, 
of operators ${\cal O}$ between generating functions $g$ of Eq. 
(\protect\ref{gfn}). Here we take all vectors $d$-dimensional. 
$\tilde{w}\vec{\bfi r}$ is a short-hand notation for 
$\sum_{i=1}^{3N} w_i\vec{ r}_i$. 
$B=A+A'$, $\vec{\bfi v}=\vec{\bfi s}+\vec{\bfi s}{\,'}$. $\vec{\bfi y}=A'B^{-1}\vec{\bfi s}
-AB^{-1}\vec{\bfi s}{\,'}$ where the elements of $w_i$ from $1$ to $N$ are repeated for $N+1$ to $2N$ and $2N+1$ to $3N$. $P$ is a permutation operator and the matrix $T_P$ 
is defined by $P\vec{\bfi r}=T_P\vec{\bfi r}$. Additionally, we define both $Q$ and $\Lambda$ as symmetric matrices.}
\label{tab7.1}
\halign{\strut #\hfill\ & #\hfill\cr
\noalign{\hrule\smallskip}
${\cal O}$ & ${\cal M}$  \cr
\noalign{\smallskip\hrule\smallskip}
 1 & ${\cal M}_0\equiv\left({\frac{(2\pi)^{n}} { {\rm det} B }}\right)^{\frac{1}{2}}
{\rm exp}({\frac{1}{2}}{{\vec{\bfi v}}} B^{-1} \vec{\bfi v})$ \cr

${\tilde{w}}\vec{\bfi r}$ & ${\tilde{w}}B^{-1}\vec{\bfi v}{\cal M}_0$ \cr

${{\vec{\bfi r}}Q\vec{\bfi r}}$  & 
$\Big\{{\rm Tr}(B^{-1}Q)+{{\vec{\bfi v}}}B^{-1}QB^{-1}\vec{\bfi v}\Big\}
{\cal M}_0 $ \cr
    
${\tilde{{\zeta}}}{\bfi \pi}$     & 
$-i\hbar {\tilde{{\zeta}}}\vec{\bfi y}{\cal M}_0 $ \cr
(${\bfi \pi}_j=-i\hbar \frac{\partial}{\partial \vec{\bfi r}_j}$) &  \cr
${\widetilde{{\bfi \pi}}\Lambda{\bfi \pi}}$    & 
$\hbar^2\Big\{{\rm Tr}(AB^{-1}A'\Lambda )-{\vec{\bfi y}}\Lambda\vec{\bfi y}
\Big\}{\cal M}_0 $ \cr

$\delta({\tilde{w}}\vec{\bfi r}-{\bfi r})$ & ${\cal M}_1\equiv
(2\pi{\tilde{w}}B^{-1}w)^{-{\frac{1}{2}}}$ \cr
   & $\quad \quad \times \,\,{\rm exp}\Big\{-\frac{1}{2{\tilde{w}}B^{-1}w}
({\bfi r}-{\tilde{w}}B^{-1}\vec{\bfi v})^2\Big\}{\cal M}_0 $ \cr
 \cr
$P$  &  $\langle g(\vec{\bfi s}'; A',\vec{\bfi r})| g({\widetilde{T_P}}\vec{\bfi s}; 
{\widetilde{T_P}}AT_P,\vec{\bfi r})\rangle$ \cr
\noalign{\smallskip\hrule}\cr}
\end{table}
\end{widetext}

\section{Matrix elements of potentials}
\label{potmatrix}
Analytical integration over $\vec r$ in Eq. \eqref{potme} for certain potentials is possible. 
These are listed in this Appendix.

\subsection{Gaussian potential} 
The Gaussian potential operator is given by
\begin{equation}
    V(r)={\rm e}^{-\mu {\vec r}^{\,2}}.
\end{equation}
The matrix element of the Gaussian
potential is obtained with the use of Eq. \eqref{potme} by
\begin{widetext}
\begin{eqnarray}
\langle\Psi_i\vert V\vert\Psi_j\rangle&=&
{\frac{1}{ ({\rm det} B)^{1/2} }} \langle\Psi_i\vert\Psi_j\rangle
{\frac{1} {({\rm det} (B^{-1}+2\mu I))^{1/2}}}
{\rm exp}\Big(-{\frac{1}{2}}{\vec {b}} B^{-1} \vec{b}
+{\frac{1}{2}}(B^{-1}{\vec {b}}) (B^{-1}+2\mu I)^{-1} (B^{-1}\vec{b})\Big)\nonumber\\
&=&
\left({\frac{{\rm det} B^{-1}} { {\rm det} (B^{-1}+2\mu I)}}\right)^{1/2}
{\rm exp}\Big(
-\mu (B^{-1}{\vec {b}}) (B^{-1}+2\mu I)^{-1}\vec{b}
\Big) \langle\Psi_i\vert\Psi_j\rangle,
\end{eqnarray}
\end{widetext}
where substituting $\mu=0$  gives back the overlap matrix as expected.\\ 

\subsection{Harmonic oscillator}
The harmonic oscillator operator is given by
\begin{equation}
    V(r)={\vec r}^{\,2},
\end{equation}
and its matrix element is given by
\begin{equation}
\langle\Psi_i\vert V\vert\Psi_j\rangle=
{\rm Tr}(B)+\vec{b}^{\,2}.
\end{equation}

\subsection{Coulomb Potential}

Using the definition
\begin{equation}
    \frac{1}{r}=\frac{2}{\sqrt{\pi}}\int_0^{\infty} \, e^{-r^2\rho^2}d\rho,
\end{equation} 
\begin{widetext}
we can evaluate the matrix element of the Coulomb potential
\begin{eqnarray}
\label{c6}
\langle \Psi|\frac{1}{|\vec{r_i}-\vec{r_j}|}|\Psi'\rangle
&=&\frac{2}{\sqrt{\pi}}\int_0^{\infty} \, 
\langle \Psi|e^{-\rho^2 (w^{ij}\vec{\bm r})^2}|\Psi'\rangle d\rho\notag \\
&=&\frac{2}{\sqrt{\pi}} \frac{1}{(2\pi)^{3/2}({\rm det}B)^{1/2}}\langle \Psi|\Psi'\rangle \int_0^{\infty}d\rho \, \int\, \exp\Big(-\rho^2 {\vec {{r}}}^{\,2} 
-\frac{1}{2}(\vec{r}-\vec {{b}})B^{-1}(\vec {{r}}-\vec {b})\Big) d\vec{ r}.
\end{eqnarray}
The integration over $\vec r$ can be done by diagonalizing $3 \times 3$ matrix
$ B^{-1}$: 
\begin{equation}
\rho^2 {\vec {r}}^{\,2} 
+\frac{1}{2}(\vec{r}-\vec {{b}})B^{-1}(\vec{r}-\vec {{b}}) = \sum_{i=1}^3\Big[(\frac{1}{2}\gamma_i +\rho^2)t_i^2 - \mu_i t_i\Big]+\frac{1}{2}\vec {b} B^{-1}\vec {b},
\end{equation}
where $\gamma_i$ is the eigenvalue of $B^{-1}$ and $t_i$ is the corresponding 
eigenvector and $\mu_i$ is easily determined. Then the integral in Eq. \eqref{c6} is 
\begin{eqnarray}
&&\int \, \exp\Big(-\rho^2 {\vec {r}}^{\,2} 
-\frac{1}{2}(\vec{r}-\vec {{b}})B^{-1}(\vec{r}-\vec {{b}})\Big)d\vec{r}\notag \\
&&=e^{-\frac{1}{2}\vec {{b}} B^{-1}\vec {{b}}} \prod_{i=1}^3 \int_{-\infty}^{+\infty} 
\exp\big(- (\frac{1}{2}\gamma_i +\rho^2)t_i^2 +\mu_i t_i \big)dt_i
= 2^3 e^{-\frac{1}{2}\vec {{b}} B^{-1}\vec {{b}}} \prod_{i=1}^3   
\int_0^{+\infty} \,
e^{- (\frac{1}{2}\gamma_i +\rho^2)t_i^2}\cosh (\mu_i t_i)dt_i.
\label{C8}
\end{eqnarray}
We can perform the integration over $t_i$ and $\rho$ using 
\begin{equation}
\int_0^{+\infty} \,
e^{- a^2 x^2}\cosh (b x)dx = \frac{\sqrt{\pi}}{2a} \, e^{\frac{b^2}{4a^2}},  \ \ \ a>0,
\end{equation}
which when used to evaluate Eq. \eqref{C8} leads to
\begin{equation}
    \int \, \exp\Big(-\rho^2 {\vec{r}}^{\,2} 
-\frac{1}{2}(\vec{r}-\vec {{b}})B^{-1}(\vec{r}-\vec {{b}})\Big)d\vec{r}
=\pi^{\frac{3}{2}} \, e^{-\frac{1}{2}\vec {b} B^{-1}\vec {b}} \prod_{i=1}^3 (\rho^2+\frac{1}{2}\gamma_i)^{-\frac{1}{2}}\, e^{\frac{\mu_i^2}{4(\rho^2+\frac{1}{2}\gamma_i)}}.
\end{equation}

By changing $\rho$ to $t$ by (see Ref. \cite{suzuki2008})
\begin{equation}
    \rho=\sqrt{a}\frac{t}{\sqrt{1-t^2}},
\end{equation}
$\rho$ integration in Eq. \eqref{c6} reduces to a general form
\begin{equation}
    \int_0^{\infty} \prod_{i=1}^3 (\rho^2+a_i)^{-\frac{1}{2}}\,e^{\frac{b_i}{\rho^2+a_i}}d\rho=\sqrt{a}\int_0^1 \prod_{i=1}^3 
\big[a_i+(a-a_i)t^2\big]^{-\frac{1}{2}} e^{\frac{b_i(1-t^2)}{a_i+(a-a_i)t^2}}dt.
\end{equation}
It is clear that the integral reduces to the error function when $a_i$ is independent of $i$ and $a$ is set to that
common value of $a_i$. Even though $a_i$  differs from each other, by choosing $a$ equal to, say, the maximum
of $a_1, a_2, a_3$, the above integrand is a smooth function of $t$ in $[0,1]$ and therefore the integral can be
accurately evaluated numerically.
\end{widetext}

\section{Two-particle  probability}
\label{2prob}
We want to calculate the probability of finding a particle in position
$\vec{r}$ and a second one at $\vec{r}'$ defined as
\begin{equation}
P(\vec{r},\vec{r}')=\sum_{i<j}\langle \Psi_i|\delta(\tilde{
w}^i\vec{\bfi {r}}-\vec{r})\delta(\tilde{w}^j\vec{\bfi{
r}}-\vec{r'})|\Psi_j\rangle ,
\end{equation}
where $\tilde{w}^j$ is defined for $U$ in the main part. See Eq. \eqref{externalpot}. Using again the Fourier representation of
the $\delta$ function
\begin{equation}
    \delta(\tilde{w}\vec{\bfi r}-{\vec r})={\frac{1}  {(2\pi)^3}}\int {\rm e}^{i\vec{k}
    (\tilde{w}\vec{\bfi r}-{\vec r})} d{\vec k} ,
\end{equation}
\begin{widetext}
we want to evaluate the integral 
\begin{equation}
\langle \Psi_i|\delta(\tilde{ w}^i\vec{\bfi {r}}-\vec{r})\delta(\tilde{w}^j\vec{\bfi{ r}}-\vec{r'})|\Psi_j\rangle=\frac{1}{(2\pi)^6}\iint  e^{-i\vec{k}^i\cdot \vec{r} -i \vec{k}^j\cdot \vec{r'} }
     \label{corrf}
\langle \Psi_i|e^{i\vec{k}^i\cdot \tilde{ w}^i\vec{\bfi {r}}+i{\vec{k}^j}\cdot \tilde{w}^j\vec{\bfi {r}}}|\Psi_j\rangle \,d\vec{k}^i\,d\vec{k}^j\,. \\
 \end{equation}
We define a 3N-dimensional vector $\vec{\mathbf{K}}_j$ as
 \begin{eqnarray}
  \vec{k}^j\cdot \tilde{w}^j\vec{\bfi{r}}&=&\sum_{i=1}^N w_i^j
  \vec{k}^j\cdot\vec{r}_i
=\sum_{i=1}^N w_i^j(k_1^jx_i+k_2^jy_{i}+k_3^jz_{i})=\sum_{i=1}^N
w_i^j(k_1^jr_i+k_2^jr_{N+i}+k_3^jr_{2N+i})\notag \\ \nonumber
&=&(k_1^j \tilde{w}^j, k_2^j\tilde{w}^j, k_3^j \tilde{w}^j )\vec{\bfi{r}}
\equiv \vec{{\bfi{K}}}_j \vec{\bfi{r}}, \\
 \end{eqnarray}
where
\begin{equation}
\vec{\mathbf{K}}_j=\left(\begin{array}{c}
k_1^jw_1^j\\
k_1^jw_2^j\\
\vdots \\
k_1^jw_N^j \\
k_2^jw_1^j \\
\vdots \\
k_2^jw_N^j \\
k_3^jw_1^j \\
\vdots \\
k_3^jw_N^j\\
\end{array}
\right),
\end{equation}
The matrix element,  $\langle \Psi_i|e^{i\vec{\mathbf{K}}_{i}\cdot 
\vec{\bfi{r}}+i{\vec{\mathbf{K}}_{j}}\cdot \vec{\bfi{r}}}|\Psi_j\rangle$, is found to be
\begin{equation}
\langle \Psi_i|e^{i\vec{\mathbf{K}}_{i}\cdot
\vec{\bfi{r}}+i{\vec{\mathbf{K}}_{j}}\cdot\vec{\bfi{r}}}|\Psi_j\rangle
 =\int 
 e^{-\frac{1}{2}{\vec{\bfi{r}} A\vec{\bfi{r}}+\vec{\bfi{s}}
 \vec{\bfi{r}}+ i \vec{\mathbf{K}}_{i} \vec{\bfi{r}} +
i\vec{\mathbf{K}}_{j} \vec{\bfi{r}}}}\,d{\vec{\bm r}}
=\frac{(2\pi)^{\frac{3N}{2}}}{({\rm det}A)^{\frac{1}{2}}}
\,e^{\frac{1}{2}{\vec{\mathbf{Q}}A^{-1}\vec{\mathbf{Q}}}},\\
\end{equation}
\end{widetext}
where $\vec{\mathbf{Q}}$ is $3N$-dimensional column vector defined by
\begin{equation}
\vec{\mathbf{Q}}=i\vec{\mathbf{K}}_{i}+i\vec{\mathbf{K}}_{j}+\vec{\mathbf{
s}}\equiv i \vec{\mathbf{K}}_{ij}+\vec{\bfi{s}},\\
\end{equation}
which gives
\begin{equation}
{\vec{\mathbf{Q}}}A^{-1}\vec{\mathbf{Q}}=-{\vec{\mathbf{K}}_{ij}}A^{-1}
\vec{\mathbf{K}}_{ij} 
+2i{\vec{\bfi{s}}} A^{-1} \vec{\mathbf{K}}_{ij} 
+{\vec{\bfi{s}}}A^{-1}\vec{\bfi{s}}.\\
\end{equation}
Substituting this result into Eq. \eqref{corrf} leads to
\begin{widetext}
\begin{equation}
    P(\vec{r},\vec{r'})
=\langle \Psi_i|\Psi_j \rangle \frac{1}{(2\pi)^6}\iint  \exp 
\left[-i\vec{k}^i\cdot \vec{r} -i \vec{k}^j\cdot
\vec{r'}-\frac{1}{2}{\vec{\mathbf{K}}_{ij}}A^{-1} \vec{\mathbf{K}}_{ij} 
+i{\vec{\bfi{s}}} A^{-1} \vec{\mathbf{K}}_{ij}
\right]\,d\vec{k}^i\,d\vec{k}^j,
\end{equation}
\end{widetext}
where $\langle \Psi_i|\Psi_j \rangle$ is given by Eq. \eqref{overlap}. The exponent of the integrand can be expressed as
\begin{widetext}
\begin{eqnarray}
    &&-i\vec{k}\cdot \vec{r} -i \vec{k}^j\cdot
    \vec{r'}-\frac{1}{2}{\vec{\mathbf{K}}_{ij}}A^{-1} \vec{\mathbf{K}}_{ij} 
+i{\vec{\bfi{s}}} A^{-1} \vec{\mathbf{K}}_{ij}\notag \\
&&=-\frac{1}{2}\sum_{\alpha,\beta=1}^3 \Big(\tilde{ w}^iA^{-1}_{\ \alpha\beta}w^ik_\alpha^i k_\beta^i+\tilde{ w}^iA^{-1}_{\ \alpha\beta}w^j k_\alpha^i k_\beta^j
+\tilde{ w}^jA^{-1}_{\ \alpha\beta}w^i k_\alpha ^jk_\beta^i +\tilde{ w}^jA^{-1}_{\ \alpha\beta}w^jk_\alpha ^jk_\beta^j\Big)\notag \\
&&\quad +i
\sum_{\alpha,\beta=1}^3\Big({\bm s_\alpha }A^{-1}_{\ \alpha\beta}w^i k_\beta^i+{\bm s_\alpha }A^{-1}_{\ \alpha\beta}w^j k_\beta^j\Big)-i\sum_{\alpha=1}^3(r_\alpha k_\alpha^i +r_\alpha 'k_\alpha ^j),
\end{eqnarray}
If we define the matrix Q as a 6 $\times$ 6 symmetric matrix given by\\
\begin{align}
Q=\left(
\begin{array}{cccccc}
\tilde{w^i}A^{-1}_{\ 11}w^i & \tilde{w^i}A^{-1}_{\ 12}w^i & \tilde{w^i}A^{-1}_{\ 13}w^i & \tilde{w^i}A^{-1}_{\ 11}w^j & \tilde{w^i}A^{-1}_{\ 12}w^j & \tilde{w^i}A^{-1}_{\ 13}w^j \\
\tilde{w^i}A^{-1}_{\ 21}w^i & \tilde{w^i}A^{-1}_{\ 22}w^i & \tilde{w^i}A^{-1}_{\ 23}w^i & \tilde{w^i}A^{-1}_{\ 21}w^j & \tilde{w^i}A^{-1}_{\ 22}w^j & \tilde{w^i}A^{-1}_{\ 23}w^j \\
\tilde{w^i}A^{-1}_{\ 31}w^i & \tilde{w^i}A^{-1}_{\ 32}w^i & \tilde{w^i}A^{-1}_{\ 33}w^i & \tilde{w^i}A^{-1}_{\ 31}w^j & \tilde{w^i}A^{-1}_{\ 32}w^j & \tilde{w^i}A^{-1}_{\ 33}w^j \\
\tilde{w^j}A^{-1}_{\ 11}w^i & \tilde{w^j}A^{-1}_{\ 12}w^i & \tilde{w^j}A^{-1}_{\ 13}w^i & \tilde{w^j}A^{-1}_{\ 11}w^j & \tilde{w^j}A^{-1}_{\ 12}w^j & \tilde{w^j}A^{-1}_{\ 13}w^j \\
\tilde{w^j}A^{-1}_{\ 21}w^i & \tilde{w^j}A^{-1}_{\ 22}w^i & \tilde{w^j}A^{-1}_{\ 23}w^i & \tilde{w^j}A^{-1}_{\ 21}w^j & \tilde{w^j}A^{-1}_{\ 22}w^j & \tilde{w^j}A^{-1}_{\ 23}w^j \\
\tilde{w^j}A^{-1}_{\ 31}w^i & \tilde{w^j}A^{-1}_{\ 32}w^i & \tilde{w^j}A^{-1}_{\ 33}w^i & \tilde{w^j}A^{-1}_{\ 31}w^j & \tilde{w^j}A^{-1}_{\ 32}w^j & \tilde{w^j}A^{-1}_{\ 33}w^j \\
\end{array}
\right),
\end{align}
and $\bm V$ as a 6-dimensional column vector defined by
\begin{align}
\bm V=\sum_{\alpha=1}^3\left(
\begin{array}{c}
\tilde{w^i}A^{-1}_{\ 1\alpha}\vec{\bm s_\alpha} \\
\tilde{w^i}A^{-1}_{\ 2\alpha}\vec{\bm s_\alpha} \\
\tilde{w^i}A^{-1}_{\ 3\alpha}\vec{\bm s_\alpha} \\
\tilde{w^j}A^{-1}_{\ 1\alpha}\vec{\bm s_\alpha} \\
\tilde{w^j}A^{-1}_{\ 2\alpha}\vec{\bm s_\alpha}\\
\tilde{w^j}A^{-1}_{\ 3\alpha}\vec{\bm s_\alpha} \\
\end{array}
\right)-
\left(
\begin{array}{c}
r_1 \\
r_2 \\
r_3 \\
r_1' \\
r_2'\\
r_3'\\
\end{array}
\right).
\end{align}
We can carry out the integration over $\vec{k},\vec{k}'$ and express the probability density as 
\begin{equation}
    P(\vec{r},\vec{r'})
=\langle \Psi_i|\Psi_j\rangle \frac{1}{(2\pi)^6} \int 
e^{-\frac{1}{2}\tilde{\bm t}Q\bm t + i \widetilde{\bm V}\bm t} d{\bm t}=
\langle \Psi_i|\Psi_j \rangle \frac{1}{(2\pi)^6}\frac{(2\pi)^3}{({\rm det}Q)^{\frac{1}{2}}}e^{-\frac{1}{2}\widetilde{\bm V}Q^{-1}\bm V}.
\end{equation}
\end{widetext}
To check this expession we integrate it over $\vec{r},\vec{r}'$. Defining the six dimensional column vectors $\vec{c}$ and $\vec{\xi}$ as
\begin{align}
\vec{c}=\sum_{\alpha=1}^3\left(
\begin{array}{c}
\tilde{w^i}A^{-1}_{\ 1\alpha}\vec{\bm s_\alpha} \\
\tilde{w^i}A^{-1}_{\ 2\alpha}\vec{\bm s_\alpha} \\
\tilde{w^i}A^{-1}_{\ 3\alpha}\vec{\bm s_\alpha} \\
\tilde{w^j}A^{-1}_{\ 1\alpha}\vec{\bm s_\alpha} \\
\tilde{w^j}A^{-1}_{\ 2\alpha}\vec{\bm s_\alpha}\\
\tilde{w^j}A^{-1}_{\ 3\alpha}\vec{\bm s_\alpha} \\
\end{array}
\right),
\end{align}
and 
\begin{align}
\vec{\xi} =\left(
\begin{array}{c}
\vec r_1 \\
\vec r_2 \\
\vec r_3 \\
\vec r_1\,' \\
\vec r_2\,'\\
\vec r_3\,'\\
\end{array}
\right).
\end{align}
\begin{widetext}
we can  integrate  the probability density over $r$ and $r'$ as\\
\begin{equation}
\langle \Psi_i|\Psi_j \rangle \frac{1}{(2\pi)^6}\frac{(2\pi)^3}{({\rm det}Q)^{\frac{1}{2}}}\iint{e^{-\frac{1}{2}\widetilde{\bm V}Q^{-1}\bm V}d\vec{r}d\vec{r'}}=\langle \Psi_i|\Psi_j \rangle \frac{1}{(2\pi)^6}\frac{(2\pi)^3}{({\rm det}Q)^{\frac{1}{2}}}\int{e^{-\frac{1}{2}{\vec c}Q^{-1}{\vec c}-\frac{1}{2}{\vec \xi}Q^{-1}{\vec \xi} +{\vec \xi}Q^{-1}{\vec c} }d\vec{\xi}},
\end{equation}
where the integral evaluates to 
\begin{equation}
    \int{e^{-\frac{1}{2}{\vec c}Q^{-1}{\vec c}-\frac{1}{2}{\vec \xi}Q^{-1}{\vec \xi} +{\vec \xi}Q^{-1}{\vec c} }d\vec{\xi}}=\frac{(2\pi)^3}{({\rm det}Q^{-1})^{\frac{1}{2}}}{e^{-\frac{1}{2}{\vec c}Q^{-1}{\vec c}}}{e^{\frac{1}{2}{\vec c}Q^{-1}{\vec c}}} = \frac{(2\pi)^3}{({\rm det}Q^{-1})^{\frac{1}{2}}},
\end{equation}
and we get back the overlap.
\end{widetext}

\section{Generalization of dipole self-interaction}
\label{gendip}
Eq. \eqref{self} can be generalized straightforwardly to multiphoton modes. By multiplying the kinetic operator with DECG exponential in each photon space, we can remove the quadratic in an analogous way.
\begin{widetext}
\begin{equation}
-\frac{1}{2}\sum_{n=1}^{N_p}\sum_{i=1}^{3N}\left(\frac{\partial^{2}}{\partial r_{i}^{2}}\right) \exp \left(\alpha_n(\vec{\lambda}_n \cdot \vec{D})^{2}\right)=
-\sum_{n=1}^{N_p}\left({\frac{1} {4\alpha_n}}+{\frac{1}{2}}(\vec{\lambda}_n \cdot \vec{D})^{2}\right) \exp \left(\alpha_n(\vec{\lambda}_n \cdot \vec{D})^{2}\right).
\end{equation}
\end{widetext}
where
\begin{equation}
    \alpha_n=\frac{1}{2\sqrt{\sum_{i=1}^{N}q_{i}^2}\,\lambda_n}
\end{equation}

\begin{acknowledgments}
This work has been supported by the National Science
Foundation (NSF) under Grant No. IRES 1826917.
\end{acknowledgments}

\par\noindent
{\bf DATA AVAILABILITY}
\par\noindent
Data available on request from the authors.
\newpage

\end{document}